\input{psfig}
\documentstyle [12pt,twoside,epsf]{article}

\newcommand{\be}{\begin{equation}}
\newcommand{\ba}{\begin{eqnarray}}
\newcommand{\ee}{\end{equation}}
\newcommand{\ea}{\end{eqnarray}}

\def\sh{\mbox{sh}}
\def\ch{\mbox{ch}}

\def\b{\beta}

\def\d{\delta}
\def\e{\epsilon}

\def\l{\lambda}

\def\o{\omega}
\def\p{\pi}

\def\r{\rho}

\def\t{\tau}

\def\G{\Gamma}

\def\rarr{\rightarrow}

\def\bo{{\raise.15ex\hbox{\large$\Box$}}}
\def\bob{{\lower.2ex\hbox{\large$\Box$}}}
\def\pa{\partial}

\def\bra#1{\left\langle #1\right|}
\def\ket#1{\left| #1\right\rangle}

\catcode`@=11
\def\underline#1{\relax\ifmmode\@@underline#1\else
        $\@@underline{\hbox{#1}}$\relax\fi}
\catcode`@=12
 
\begin{document}

\begin{titlepage}
\rightline{CERN-TH.7432/94}
\rightline{LA-UR-94-3410}
\rightline{August 22, 1994} \vskip .1in

\centerline{\large{\bf Finite Energy Instantons in the $O(3)$
Non-linear Sigma Model}} 

\vspace{1cm}

\centerline{\bf Peter G. Tinyakov$^{\star}$, Emil Mottola$^{\dagger}$,
and Salman Habib$^{\dagger}$}

\vspace{.5cm}

\centerline{{\em $^*$CERN}}
\centerline{{\em European Laboratory for Particle Physics}}
\centerline{{\em CH-1211 Gen\'eve 23}}
\centerline{{\em and}}
\centerline{\em Institute for Nuclear Research of the
Russian Academy of Sciences} 
\centerline{\em 60th October Anniversary Prospect, 7a}
\centerline{\em Moscow, 117312}  

\vspace{.3cm}
 
\centerline{\em $^{\dagger}$Theoretical Division}
\centerline{\em Los Alamos National Laboratory}
\centerline{\em Los Alamos, NM 87545}

\vspace{.5cm}

\centerline{\bf Abstract}

\vspace{.5cm}

We consider winding number transitions in the two dimensional $O(3)$
non-linear sigma model, modified by a suitable conformal symmetry
breaking term. We discuss the general properties of the relevant
instanton solutions which dominate the transition amplitudes at finite
energy, and find the solutions numerically. The Euclidean period of
the solution {\em increases} with energy, contrary to the behavior
found in the abelian Higgs model or simple one dimensional systems.
This indicates that there is a sharp crossover from instanton
dominated tunneling to sphaleron dominated thermal activation at a
certain critical temperature in this model. We argue that the
electroweak theory in four dimensions should exhibit a similar
behavior.

\vfill
\noindent e-mail:\\
\noindent peter@amber.inr.free.net\\
\noindent emil@pion.lanl.gov\\
\noindent habib@predator.lanl.gov
\end{titlepage}
 
\newpage

Gauge theories of the strong and electroweak interactions are
characterized by a multiple vacuum structure. Tunneling transitions
between different vacua are responsible for physically interesting
effects, such as baryon number violation in the electroweak theory. At
zero temperature and energy these winding number transitions are
dominated by the familiar zero energy instanton solutions of the
Euclidean field equations, with vacuum boundary conditions.

At finite temperatures thermal activation over the potential barrier
separating the multiple vacua can occur in addition to quantum
tunneling. The static classical solution whose energy is equal to the
top of this barrier between neighboring vacua is the sphaleron
solution.  At sufficiently high temperatures transitions between
different winding number sectors are dominated by classical thermal
activation with a rate controlled by the energy of the sphaleron.

In simple one dimensional systems there is a smooth crossover from the
zero energy instanton dominated tunneling transition to the high
temperature sphaleron dominated regime. The corresponding classical
solutions interpolating between these two situations are known as
periodic instantons with turning points at finite Euclidean time $\b$,
which give a non-pertubative contribution to the partition function
and transition rate at temperature $\b^{-1}$. One would expect similar
considerations to apply in quantum field theory, although the
situation is much less well explored and very few classical solutions
of this kind are known.

Periodic instantons appear naturally also in the context of zero
temperature field theory, if one considers the transition probability
between different winding number sectors at fixed energy, $E$. This
probability may be expressed in the form,
\be 
P(E)= \sum_{i,f} |\bra{f}{\cal S} {\cal P}_E\ket{i}|^2 \; ,
\label{P[E]} 
\ee 
where ${\cal S}$ is the ${\cal S}$-matrix, ${\cal P}_E$ is a projector
onto energy $E$ and the initial and final states $\ket{i}$ and
$\bra{f}$ lie in different winding sectors. The path integral for
$P(E)$ is saturated by a periodic instanton solution \cite{KRTperiod},
and the probability is given, with exponential accuracy, by the exact
analog of the quantum-mechanical formula,
\be 
P(E)\sim \exp[E\b-S(\b)]\; , \label{P[E]semicl} 
\ee
where $S(\b)$ is the Euclidean action of the periodic instanton. The
Euclidean period $\b$ is related to the energy by
\be 
{dS(\b)\over d\b}=E\; .  \label{energy cond} 
\ee 
The initial and final multi-particle states can be read off from the
analytic continuation of the periodic instanton into Minkowski time at
its turning points. Hence periodic instanton solutions to the
classical Euclidean equations contain non-trivial information about
multi-particle transition amplitudes between different winding number
sectors at finite energy. It has been suggested that by suitably
modifying the boundary conditions in the complex time plane,
information about transition amplitudes from few particle initial
states to many particle final states may be obtained as well
\cite{few-many}. For these reasons it is interesting to study more
thoroughly the possible classical periodic instanton solutions in
field theories with winding number transitions.

Because of the absence of analytical methods for finding periodic
instanton solutions to the Euclidean equations in interesting field
theories, one must rely on a numerical approach. In earlier work
periodic instantons have been studied numerically in the two
dimensional abelian Higgs model \cite{MatveevJ}. In this letter we
present the results of a numerical study of periodic instanton
solutions in the two dimensional non-linear $O(3)$ sigma model, which
differs from the abelian Higgs model by the absence of a scale
parameter (before the addition of any $O(3)$ symmetry breaking). As a
result, zero energy instantons in the unbroken $O(3)$ sigma model can
have arbitrary size, exactly as is the case for unbroken $SU(2)$
gauge theory in four dimensions. As we shall see, this leads to a 
dependence of the period of the periodic instanton on energy which is 
quite different from that observed in systems with only one degree of 
freedom or the abelian Higgs model, and a sharp rather than smooth 
transition from quantum tunneling to thermal activation at a certain 
critical temperature.

The two-dimensional $O(3)$ sigma model has the Euclidean action 
\be
S_0={1\over 2g^2} \int d^2x \,(\pa_{\mu}n_a)^2\; ,
\label{action0}
\ee
where $n_a(x)$, $a=1,2,3$ are three components of a unit vector,
$n_an_a=1$. The model is well known to possess instanton solutions
\cite{sigmainst}, which may be pictured geometrically by first conformally
mapping two dimensional Euclidean space $R^2$ onto $S^2$ with all points at
infinity identified, and then taking the field configuration which
is just the identity mapping of this $S^2$ onto the $S^2$ of $n_a$
field space. Hence the zero energy instanton covers the entire sphere
pictured in Fig. 1 as Euclidean space and time vary over the infinite
range $[-\infty , +\infty]$.  Because of the conformal invariance of
the action (\ref{action0}), the zero energy instanton can have
arbitrary scale $\r$.  This implies that the energy barrier between
winding number sectors, proportional to $1/\r$, can be made
arbitrarily small and no finite energy sphaleron solution exists in
the symmetric $O(3)$ model. This is precisely as in the pure
Yang-Mills theory in four dimensions.

\begin{figure}
\centerline{\psfig{figure=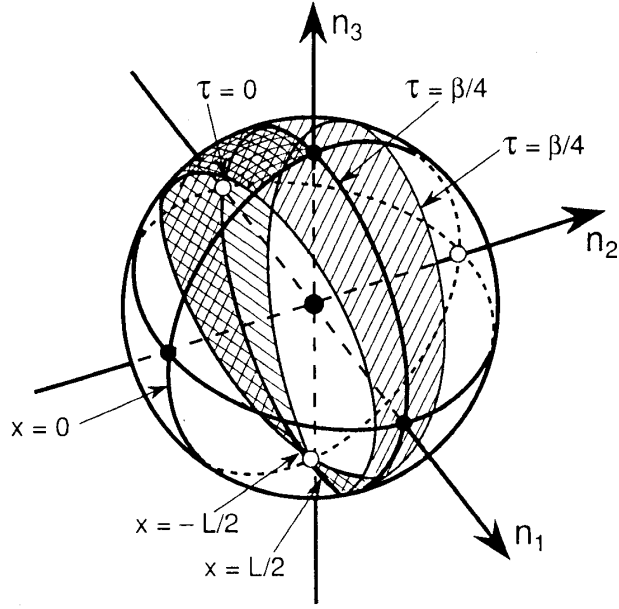,height=8cm,width=8cm}}
\caption[Figure 1] {\small{Geometric representation of the periodic
instanton.}}   
\end{figure}

In the electroweak theory conformal invariance is broken in the Higgs
sector which then makes possible the existence of a classical
sphaleron solution. In the $O(3)$ sigma model the conformal invariance
can be broken by adding to the action (\ref{action0}) the term
\be
S_1={m^2\over g^2}\int d^2x (1+n_3)\; ,
\label{action1}
\ee
which also violates the O(3) symmetry and fixes the vacuum state to be
$n^{a\,(vac)}=(0,0,-1)$. With the conformal symmetry breaking term
(\ref{action1}), the modified model with total action $S = S_0 + S_1$
possesses the classical sphaleron solution \cite{Mottola&Wipf}
\be 
n^{(sph)}_a(x)=\biggl( -2{\sh (mx)\over \ch^2(mx)},\,
0\,,-1+{2\over\ch^2(mx)} \biggr)\; , \label{sphaleron} 
\ee 
which has the energy 
\be
E_{sph}={8m\over g^2}\,. \label{sphener}
\ee
Geometrically, this solution maps the infinite spatial line
onto a great circle beginning and ending at the south pole in Fig. 1.

The spectrum of perturbations around the sphaleron has exactly one
normalizable negative mode with the eigenfunction
$u_a(x)=(0,\,\ch^{-2}(mx),\,0)$ and the eigenvalue $\omega_-^2=-3m^2$.
Because of the existence of this negative mode it is possible to
determine the form of the periodic instanton for energies just below
the sphaleron energy (\ref{sphener}). Adding a small amplitude $\e$ of the
negative mode to the static sphaleron gives a Euclidean time-dependent
periodic instanton solution,
\be
n^{(sph)}_a(x) + \e\;  {\sin}(|\o_-|\t) u_a(x) \label{nearEsph}
\ee
with an energy just below the sphaleron energy. The period of
the periodic instanton approaches the critical value
$\b_{crit}= 2 \p /|\o_-| = 2\pi/(\sqrt{3}m)\approx 3.628/m$ as 
$\e\rarr 0$ and the energy approaches the sphaleron energy, $E\rarr E_{sph}$. 

In the other limit of very low energies, $E\ll E_{sph}$, 
the periodic instanton solution can be found by considering an infinite 
chain of instanton-anti-instanton pairs on the Euclidean time axis. The
infinite chain acts as a set of images for the field configuration in
one link with fundamental period $\b$. The classical action of this
approximate solution over the fundamental period is 
\be 
S(\b) \approx {8\p\over g^2} \Bigl(1 - \pi^2{\r^2\over \b^2} -  m^2\r^2
\ln (m\r)\Bigr)\ .  
\ee 
The corrections to the action are suppressed by factors of $m^2\b^2$,
$m^2\r^2$ and $\r^2/\b^2$ which all go to zero for $E \ll E_{sph}$, as
we shall see in a moment.  Substituting this action into the
expression for the rate of tunneling transitions at fixed energy,
(\ref{P[E]semicl}) gives
\be
P(E)\sim \exp \Bigl( -{1\over g^2} W \Bigr)
\ee
with 
\be
W= 8\pi - g^2\b E - 8\pi^3{\rho^2\over \b^2} - 8\pi m^2\r^2\ln
(m\r)\  \label{W lowE}
\ee
where the values of the period $\b$ and the instanton size $\rho$ are
determined by the saddle-point equations \cite{KRTperiod}
\be
{\pa {W}\over\pa{\b}}={\pa W\over\pa{\rho}}=0.
\ee
In the limit $E\ll E_{sph}$ we find
\ba
\b(E)&\approx &{1\over m}\biggl\{ {2\pi^2\over \ln(m/g^2E)} \biggr\}^{1/2},
\label{T}\\ 
\rho(E)&\approx &{(g^2E/m)^{1/2}\over 2^{5/4}m}\biggl\{ {1 \over \ln(m/g^2E)}
\biggr\}^{3/4},         \label{rho}
\ea
which implies
\be
W(E)\approx 8\pi - {g^2E\over m} \biggl\{ {2\pi^2\over
\ln(m/g^2E)}\biggr\}^{1/2}\ , 
\ee
and justifies the neglect of the corrections to the action in the
limit of very small energies. 

This approximate behavior of the periodic instanton at low energy
leads us to observe that the conformal symmetry breaking term in the
action (\ref{action1}) serves only to fix the scale of energy and
instanton size. It does not distort the periodic instanton from the
configuration obtained by a simple linear superposition of zero energy
instanton-anti-instanton pairs of the model with unbroken symmetry.
This is possible only because $\b (E)$ goes to zero as $E \rarr 0$,
{\em i.e.},  $\b(E)$ is an {\em increasing} function of energy,
\be
{d \b (E) \over dE} > 0 \ . \label{incr}
\ee
Notice that this behavior is nevertheless consistent with the linear
superposition or dilute gas approximation because the parameter that
controls the validity of this approximation is $\r^2/\b^2$, and $\r$
goes to zero even faster than $\b$ as $E\rarr 0$.  Hence it is already
clear that the behavior of the periodic instantons in the $O(3)$ model
is very different from that of systems with only one quantum
mechanical degree of freedom or the abelian Higgs model. In those
models $\b(E) \rarr \infty$ as $E\rarr 0$, and the periodic instanton
contributes to the transition rate as the temperature $\b^{-1}$ goes
smoothly to zero \cite{Aff}.  Periodic instantons in the present model
with period going to {\em zero} as $E\rarr 0$ cannot contribute to the
low temperature transition rate, although they can contribute, and in
fact dominate transitions between {\em non}-thermal low energy states.
We shall see presently that the behavior of the period with energy
(\ref{incr}) persists up to the sphaleron energy.

At energy comparable, but not very close to the sphaleron energy, the
periodic instanton solution has to be found numerically.  For that one
must solve the Euclidean field equations
\be
-(\pa_\t^2 + \pa_x^2) n_a + m^2 \d_{a3}-\l n_a=0,
\label{eqsom}
\ee
together with the constraint,
\be
n_an_a=1\,.            \label{cont eqs}
\ee
Here, $\lambda$ is a Lagrange multiplier enforcing the constraint,
which is easily found by multiplying (\ref{eqsom}) by $n_a$ and using
(\ref{cont eqs}). The periodic instanton solution we seek has
vanishing time derivative at initial time $\t=0$, evolves to another
turning point at half-period $\b/2$ where it reflects and then returns
to its initial configuration at $\t = \b$ by simply reversing the sign
of all $\t$ derivatives. Thus we enforce the half-period boundary
conditions,
\be
{\pa n_a(\t=0,x)\over\pa \t}={\pa n_a(\t=\b/2,x)\over\pa \t}=0 \ .
\ee
This removes the time translation invariance of the solution. The
other boundary condition we require is that at spatial infinity the
solution approaches the vacuum. Since we shall work in a finite box of
length $L$ we require,
\be
n_a(\t, x=-L/2)= n_a(\t, x=L/2) = n^{(vac)}_a = (0,0,-1)
\ee
for $mL \gg 1$. 

\begin{figure}
\centerline{\psfig{figure=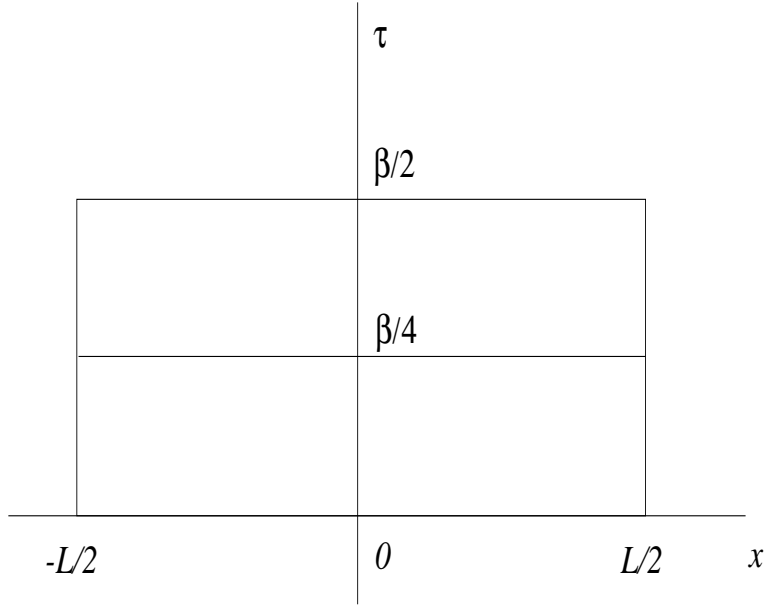,height=8cm,width=10cm}}
\caption[Figure 1] {\small{The finite rectangular region $(\tau,x)$ of
coordinate space that maps onto the shaded region of the sphere in
Fig. 1.}}    
\end{figure}

Geometrically, the periodic instanton solution obeying these boundary
conditions maps the rectangular region of $(\t,x)$ coordinate space
pictured in Fig. 2 into the shaded region of the sphere in
Fig. 1. From these figures it should be clear that the solution may be
chosen to have well-defined symmetry properties under reflection
through the lines bisecting the rectangular region of Fig. 2, so that
at $x=0$, $n_1$ vanishes while $n_2$ and $n_3$ reach extrema, and at
$\t = \b/4$, $n_2$ vanishes while $n_1$ and $n_3$ reach
extrema. Imposing these symmetries fixes completely the spatial
translational invariance and rotation invariance around the $n_3$ axis
of the solution, as well as reduces the region we must consider to
only one quarter of the full rectangle in Fig. 2. Hence we must solve
the differential eqs. (\ref{eqsom}) subject to the boundary
conditions,
\be
\begin{array}{lcl}
x=-L/2 & : & n_1=0;\;\;n_2=0;\;\;n_3=-1; \\
t=0    & : & dn_1/dt=dn_2/dt=dn_3/dt=0;\\
x=0    & : & n_1=0;\;\;dn_2/dx=dn_3/dx=0;\\
t=\b/4  & : & n_2=0;\;\;dn_1/dt=dn_3/dt=0.
\end{array}                                 \label{bc}
\ee

The lattice version of eqs. (\ref{eqsom}) can be obtained by a
variational principle starting from the discretized action. Let
$(t_i,x_j)$, where $i=0,\ldots, I$ and $j=0,\ldots, J$ , be the
coordinates of lattice sites. The lattice version of the action
(\ref{action0}-\ref{action1}) reads
\be
S=\sum\limits_{i,j} \Bigl\{{1\over 2} (n^a_{i+1,j}-n^a_{ij})^2 
{d\tilde{x}_j\over dt_i}
+{1\over 2} (n^a_{i,j+1}-n^a_{ij})^2  {d\tilde{t}_i\over dx_j}
+ m^2(1+n^3_{ij})d\tilde{t}_id\tilde{x}_j \Bigr\}
\label{action discr}
\ee
where we have used the notations $dt_i=t_{i+1}-t_i$,
$d\tilde{t}_i=(1/2)(dt_{i-1}+dt_i)$ and similarly for $dx_j$,
$d\tilde{x}_j$.  

The equations which follow from the action (\ref{action discr}) 
can be written in the form 
\be
[B^a(\delta^{ab}-n^an^b)]_{ij}= 0, \label{discr eqs}
\ee
where 
\be
B^a_{ij}\equiv - {n^a_{i+1,j}\over d\tilde{t}_idt_i}
	- {n^a_{i-1,j}\over d\tilde{t}_idt_{i-1}}
- {n^a_{i,j+1}\over d\tilde{x}_jdx_j}
- {n^a_{i,j-1}\over d\tilde{x}_jdx_{j-1}}
+ m^2\delta_{a3}. \label{Bij}
\ee
Eqs. (\ref{discr eqs}) actually contain only two independent equations
since its projection onto the vector $n^a_{ij}$ is identically
zero. Together with the constraint equation (\ref{cont eqs}) these
comprise a complete set of three independent equations associated with
each lattice site. From eqs.  (\ref{discr eqs}) it is clear that $B^a$
must be parallel to $n^a$ at each point $(i,j)$. Since $n^a$ is
normalized to unity, the eqs. (\ref{discr eqs})--(\ref{Bij}) can be
rewritten in the equivalent, symmetric form,
\be
n^a\sqrt {B^cB^c} + B^a=0,         \label{symm eqs}
\ee
for every $(i,j)$. This form is more convenient for numerical
calculations.

To obtain a  numerical solution of (\ref{symm eqs}) we use Newton's
method. That is, we \begin{enumerate}
\item choose an initial field configuration as a first guess;
\item linearize the equations in the background of the initial field;
\item solve the linearized equations to obtain an improved
configuration; and 
\item iterate until the procedure converges.
\end{enumerate} 
The choice of initial configuration is guided by the
known behavior of the periodic instanton solution near $E=0$ and
$E=E_{sph}$.  For example, for $E$ just below $E_{sph}$ an initial
configuration of the form (\ref{nearEsph}) with $\e \sim 0.5$ works
well. The convergence of the scheme is quadratic, and double precision
accuracy is typically reached in $5$ to $7$ iterations.

The solution of the linearized equations in step (3) of the algorithm
requires an inversion of the matrix of small fluctuations about the
trial configuration. Negative eigenvalues of this matrix are treated
on the same footing as positive eigenvalues, and pose no special
problem for Newton's method, which is important for the present
application.  Because of the boundary conditions which fix the
translational and rotational symmetries there are no zero eigenvalues
of the matrix near the desired solution, which otherwise would be
disastrous for this method.

For a $I\times J$ spacetime grid the matrix to be inverted has
$I^2J^2$ elements. In general, it takes of order $I^3J^3$ operations
to invert such a matrix. However, the sparseness of the matrix makes
it more efficient to follow a procedure of forward elimination and
back substitution along either the $x$ or $\t$ directions
instead. That is, starting with one edge of the region in Fig. 2, such
as $\t =0$, we solve for each $\t$ slice of the grid in terms of the
successive two $\t$ slices until the edge $\t = \b/4$ is reached,
where the boundary condition determines the unknown quantities. Then
we reverse direction and solve for the unknowns on the previous $\t$
slices successively. This allows for the matrix to be inverted in
order $IJ^3$ operations (or $I^3J$ operations if the $x$ direction is
chosen), and speeds up the algorithm considerably.  In practice, a
grid size of order $100\times 100$ can be handled on a typical
workstation, which already provides reasonable accuracy ($\sim 1\%$)
for the problem at hand.

The numerical method described works quite well and yields a periodic
instanton solution for energies in the range of $1$ to $8$ in units of
$m/g^2$. Smaller energies require finer lattices to obtain the same
accuracy, since the instanton size becomes smaller.  The results of
our calculations are summarized in Figs. 3. They were obtained
on a $40\times 40$ grid in a spatial box of size $L=8$ in units of
$m^{-1}$.

Fig. 2 illustrates a typical periodic instanton solution. As one can
see, at the turning points $\t=0$ and $\t=\b/2$ the field
configuration is non-vacuum, and the components of the vector $n^a$
cover the shaded patch of the sphere in Fig. 1.

Fig. 3a shows the dependence of the action of the periodic instanton,
in units of $1/g^2$, on its half-period, as the latter varies from
$\b/2 \sim 1.3$ to $\b/2\sim \b_{crit}/2$. The action grows smoothly
from a value close to $S_0=4\pi$ to $\b_{crit}E_{sph}/2\approx 14.5$.

Fig. 3b shows the dependence of the half-period $\b/2$ as a function
of energy.  Near the sphaleron energy ($8$ in units $m/g^2$), the
period reaches the critical period $\b_{crit}$.

Fig. 3c shows the exponent, $W=S-E\b$, in the suppression factor for
the tunneling probability $P(E)$, as a function of energy.  $W$ varies
from a value slightly below $4\pi$ at low energies to zero at $E\sim
E_{sph}$.

\begin{figure}
\centerline{\psfig{figure=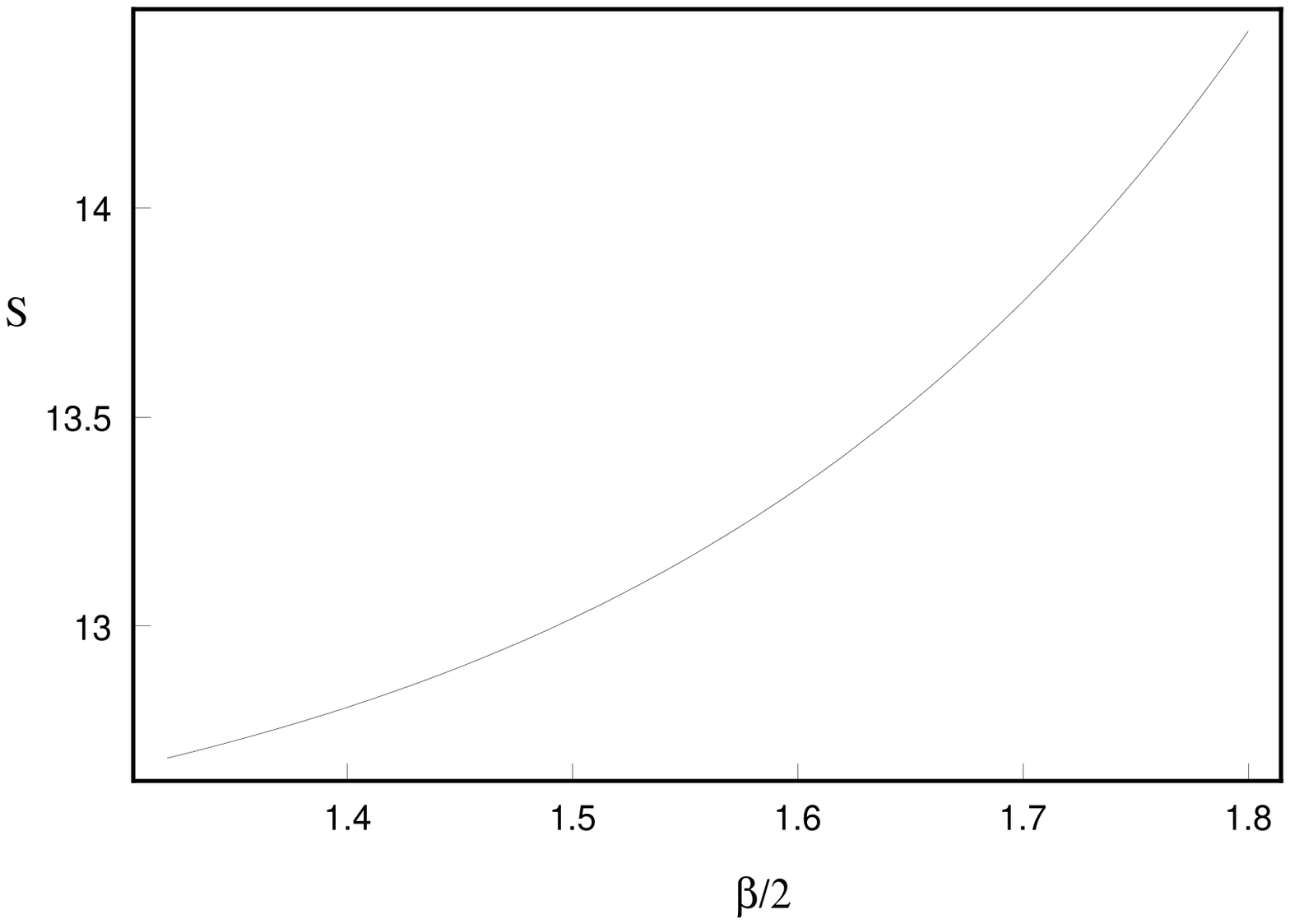,height=5cm,width=8cm}}
\caption[Figure 3] {\small{(a) Action of periodic instanton versus
half-period.}}

\vspace{1cm}

\addtocounter{figure}{-1}
\centerline{\psfig{figure=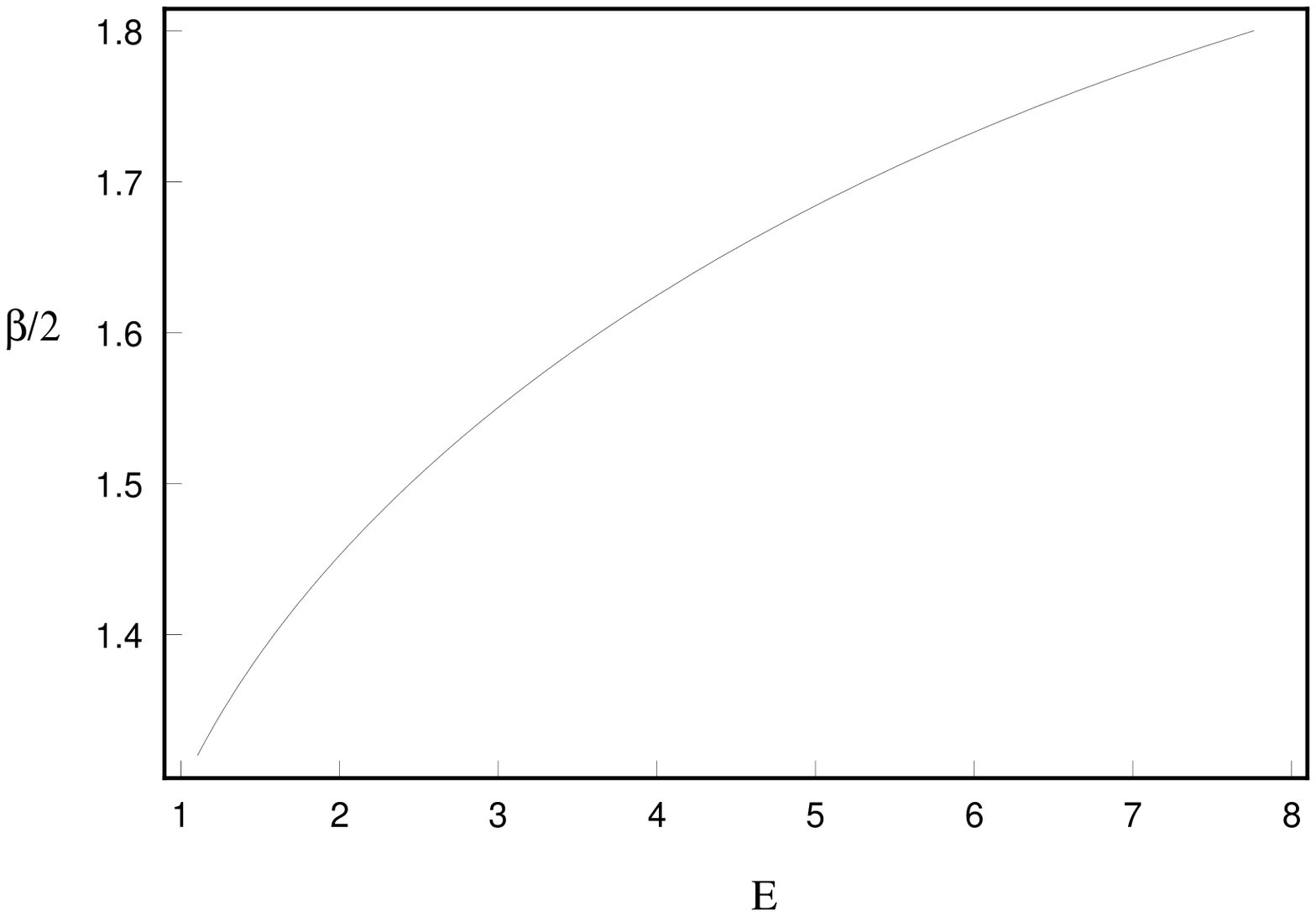,height=5cm,width=8cm}}
\caption[Figure 3] {\small{(b) Half-period versus energy for the
periodic instanton.}}    

\vspace{1cm}

\addtocounter{figure}{-1}
\centerline{\psfig{figure=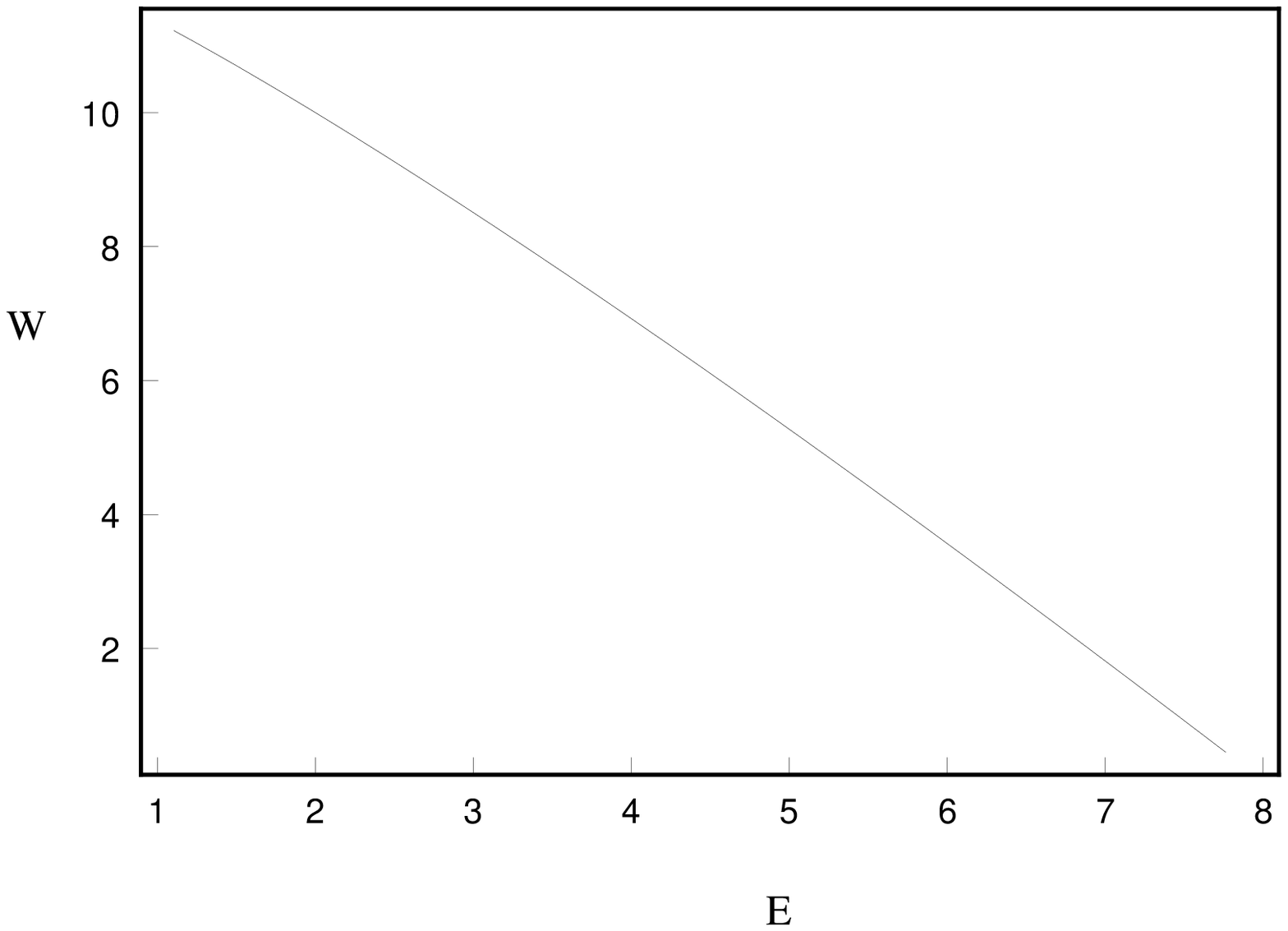,height=5cm,width=8cm}}
\caption[Figure 3] {\small{(c) The suppression factor $W$ (in $P(E)$)
as a function of energy.}}    

\end{figure}

The behavior of all quantities in these graphs is smooth, so that
greater accuracy may be obtained in a straightforward way by
increasing the number of lattice sites. 

The periodic instanton solutions we have found numerically dominate
the winding number transition rate at fixed energy.  At fixed
temperature, the transition rate is given to exponential accuracy by
\be
\G (T)=\int_0^{\infty} dE\,e^{- E/T} P(E),
\label{rate}
\ee
where $T$ is the temperature. Evaluating this integral by the method
of steepest descent (which is valid in the limit of arbitrarily weak
coupling $g^2\rarr 0$ provided $T$ is coupling independent) gives the
saddle point condition
\be
T^{-1} = -{dW\over dE} = \b (E)\ .    \label{saddle}
\ee
However, evaluating the second derivative of the exponent at this
saddle point yields
\be
-{d^2 W \over dE^2} = +{d\b (E)\over dE} > 0\ ,
\ee
which tells us that this extremum is a minimum of the exponent rather
than a maximum, for the periodic instanton solutions we have found
obeying (\ref{incr}). Hence this saddle point does not give the dominant
contribution to the rate $\G$, and the periodic solutions we have
found are {\em not} the relevant ones for finite temperature transitions.

If these periodic solutions are the {\em only} ones obeying the saddle point
periodicity condition (\ref{saddle}), then the integral in eq. (\ref{rate}) 
is saturated either at $E=0$, which implies
\[
\G(T)\sim \exp(-2 S_0)= \exp(-8\p)\; ,
\]
or at $E=E_{sph}$, where the exponent $W(E)$ reaches zero and the
exponential suppression in $P(E)$ disappears. In the latter case one
finds
\[
\G(T)\sim \exp(-E_{sph}/T)\; .
\]
The crossover between these two cases occurs when the two contributions
become equal, {\em i.e.}, at 
\[
T_{cr}={E_{sph}\over 2S_0}= {m\over \p}\; .
\]
At temperatures above $T_{cr}$ the transition rate is dominated by the
thermal activation over the top of the barrier, and the sphaleron
solution plays a dominant role. At temperatures less than $T_{cr}$,
the transition rate coincides with the zero energy tunneling rate.
At temperatures $T<T_{cr}$ there would then be no exact classical
solution which dominates the finite temperature transition rate,
and we would expect instead that the field configurations which 
saturate the rate at $T<T_{cr}$ are the same as those saturating
the zero temperature transition rate, {\em viz.}, the constrained 
instantons \cite{Affleckconstr} which are {\em not} exact solutions to 
the Euclidean field equations. Indeed, at $T=0$ the absence of any exact 
{\em real} finite action solutions immediately follows from scaling
arguments (Derrick's theorem), and at finite $T$ imposing
periodic boundary conditions at Euclidean time $T^{-1}$ does not alter
this argument as long as $T\ll m$. 

The other possibility is that the real periodic instanton solutions
we have considered in this letter are not the only solutions to the
field equations satisfying the periodicity condition (\ref{saddle}), but
that there are other solutions, perhaps complex, for which the
periodicity decreases with increasing energy, and which do dominate
the finite temperature rate $\G (T)$. Since we know that at sufficiently
low temperatures there are no other real periodic solutions than the ones we
have found, any other solutions would have to be complex, in order to
evade the scaling argument. Indeed it is interesting to
speculate that the reason the constrained instanton configurations
can dominate the zero temperature transition rate even though they
are not exact solutions is that they are a good approximation (in their
action) to some exact complex solution to the field equations, yet
to be discovered. If this speculation is correct, then we would have
to reconsider the question of whether the real periodic solutions
obeying the boundary conditions (\ref{bc}) are the correct
configurations maximizing the transition probability at fixed energy
as well.

We would like to conclude by stressing the analogy between the two
dimensional sigma model with the symmetry breaking term
(\ref{action1}) and the spontaneously broken $SU(2)$ gauge theory.  In
both cases, in the unbroken phase, instanton solutions exist and can
have arbitrary size. In the broken phase, exact real instanton solutions do
not exist, and the dominant contribution to the zero temperature
(energy) transition rate comes from approximate solutions, the
constrained instantons.  Like in the sigma model, in the spontaneously
broken $SU(2)$ gauge theory the period of the periodic instanton
increases with energy at low energies, where perturbation theory is
applicable \cite{KRTperiod}. Hence the situation is in all important
respects exactly the same as in our broken $O(3)$ sigma model, and
one may expect that the finite temperature transition rate in the $SU(2)$ 
+ Higgs system is saturated either by zero energy tunneling or by 
sphaleron activation, with a sharp crossover between these two regimes 
at temperatures of order $M_W$, or else new complex periodic solutions
to the equations of motion of the $SU(2)$- Higgs system exist which are 
still awaiting discovery.

The authors are grateful to M. Shaposhnikov and V. Rubakov for
enlightening discussions. One of us (P.T.) wishes to thank D. T. Son,
A. Kuznetsov, and S. Khlebnikov for helpful discussions, and the Theory
Division of CERN for hospitality. The work of P.T. is supported in
part by the Russian Foundation for Fundamental Research (project
93-02-3812) and by the International Science Foundation. E.M. and
S.H. acknowledge support from the United States Department of
Energy. This research was performed in part using the resources
located at the Advanced Computing Laboratory, Los Alamos National
Laboratory.

\end{document}